\documentclass[onecolumn,secnumarabic,aps,prd,notitlepage]{revtex4-1}
\usepackage{amsmath}
\usepackage{amssymb}
\usepackage[dvipdfm]{epsfig}
\usepackage{graphicx}
\usepackage{natbib}

\begin{document}

\title{Modeling and Scaling of Hysteresis in Magnetic Materials. Frequency, Pik of Induction and Temperature Dependance }  

\author{Krzysztof Z. Sokalski}

\affiliation{Institute of Computer Science, 
Cz\c{e}stochowa University of Technology, 
Al. Armii Krajowej 17, 42-200 Cz\c{e}stochowa, Poland}

%\keywords{ Mathematical model of magnetization, hysteresis loops,scaling}

%\pacs{75.50.-y, 75.60.-d, 89.75.Da}

%\date{}
%\maketitle
\begin{abstract} Recently introduced model of  magnetic hysteresis was extended into set of the following features: frequency, pick of induction and temperature of specimen. Group theoretical classification of hysteresis loops' sets is presented. An effect analogous to the Zeeman splitting has been revealed in the set of the all hysteresis loops.

\end{abstract}

\maketitle

\section{Introduction}\label{I}
  Recently new mathematical model of hysteresis loop has been derived \citep{bib:sok0}.This model consists in extantion of tanh($\cdot$) by extanding the base of exp function into an arbitrary positive number.
  \begin{equation}
  	\label{tynh1}
  	tanH(a,b,c,d|x)=\frac{a^{x}-b^{-x}}{c^{x}+d^{-x}},
  \end{equation}
  where $a,b,c,d$ are arbitrary positive numbers. In this paper we consider only a symmetric model: $a=b=c=d$, which lead to the following forms for the apper and and the lower sections of hysteresis loop as well as for the initial magnetization curve:
  \begin{equation}
  	\label{major}
  	M_{F}(X)=M_{0}F(X,\theta); \hspace{2mm}M_{G}(X)=M_{0}G(X,\theta),
  \end{equation}
  where
  \begin{eqnarray}
  	F(X,\theta)=\frac {{a}^{X+\theta}-{b}^{-X-\theta}}{{c}^{X+\theta}+{d}^{-X-\theta}}\label{d},\\
  	G(X,\theta)=\frac {{a}^{X-\theta}-{b}^{-X+\theta}}{{c}^{X-\theta}+{d}^{-X+\theta}}\label{g}.
  \end{eqnarray}
  $\theta$ is a model parameter depending on $X_{max}$. 
  \begin{equation}
  	\label{tynh2}
  	M_{P}(X)=M_{0}\,P(X,\epsilon); \hspace{2mm}X\in[0,X_{max}],
  \end{equation}
  where, $M_{0}$ is magnetization corresponding to saturation expressed in Tesla: $[$T$]$,  $X\in \{0,X_{max}\}$, $X=\frac{H}{h}$ where $H$ is magnetic field, $h$ is a parameter of the magnetic field dimension $[$A m$^{-1}]$ to be determined. $X_{max}$ is determined by an uncertainity relation \citep{bib:sok0}. $P(X,\epsilon)$ function is of the following form: 
  \begin{equation}
  	\label{primary}
  	P(X,\epsilon)=\frac {{a}^{X-\epsilon}-{b}^{-X+\epsilon}}{{c}^{X-\epsilon}+{d}^{-X+\epsilon}},
  \end{equation}
  where $\epsilon$ is  modeling parameter of the order $\theta/2$.

   In the presented paper we extend this approach into dependence of magnetization on the frequency $f$ and the induction's pick $B_{m}$ of periodic exciting wave, as well as the temperature $T$ of specimens. The $f$ dependence of hysteresis loop has been derived by Jiles in \citep{bib:JILES}.  
   His model provides a relatively simple way of predicting changes in the hysteresis curves as a result of eddy currents
   at different frequencies, however this model requires solutions of Ordinary Differential Equation. In our approach we apply phenomenological model of the hysteresis loop possessing the free parameter $a$ which will be widen to a positively determined function of $f$. Recently relevant progress has been achieved in subject of temperature dependence of the hysteresis loop \citep{bib:sutor}. The presented model consists mainly of a
   Preisach operator with a continuous Preisach weight function which is written down as a function of temperature.

  Very often, an investigator or a designer of magnetic materials is interested in dependence of hysteresis on the three parameters $f,B_{m},T$ together. Therefore
   we assume that the base $a$ is a positive determined function of the three variables: $a=a(f,B_{m},T)$.
   Crucial role in the solution of the mentioned above problems will play scaling \citep{bib:BAR},\citep{bib:SOK1},\citep{bib:Rusz}. The   reasons for creation such phenomenological models are automated, design, and the convenience as well as computational efficiency \citep{bib:reason}. 
   \section{Widening of Hysteresis' Loop Scaling on $f$, $B_{m}$ and $T$ Parameter' Space}
    Scaling of magnetization hysteresis loop introduced in \citep{bib:sok0} represents a relation between the two loops.
  
%607170459
Let the upper section of hysteresis loop and the lower one are of the forms given by (\ref{major}),(\ref{d}) and (\ref{g}), respectively. Let us consider symmetric case defined by $a=b=c=d$. Then the considered loop model is invariant, both with respect to scaling and gauge transformation represented by $\lambda$ and $\chi$ parameters, respectively. The complete transformation has to be performed for the both sections of the hysteresis loop \citep{bib:sok0}.  

 Let us assume that there exist real numbers $\{\alpha,\nu \}\in \mathbf{R}^{2}$ such that $\forall{\lambda}\in \mathbf{R}_{+}$ and $\forall{\chi}\in \mathbf{R}$ the following relations hold:
\begin{eqnarray}
\frac{M_{F}(X)}{M_{0}}\lambda^{\nu}=\frac{(\lambda^{\alpha}a)^{X+\theta}a^{\chi}-(\lambda^{\alpha}a)^{-X-\theta}a^{-\chi}}{(\lambda^{\alpha}a)^{X+\theta}a^{\chi}+(\lambda^{\alpha}a)^{-X-\theta}a^{-\chi}},\nonumber\\
\frac{M_{G}(X)}{M_{0}}\lambda^{\nu}=\frac{(\lambda^{\alpha}a)^{X-\theta}a^{\chi}-(\lambda^{\alpha}a)^{-X+\theta}a^{-\chi}}{(\lambda^{\alpha}a)^{X-\theta}a^{\chi}+(\lambda^{\alpha}a)^{-X+\theta}a^{-\chi}}.\label{scaling2}
\end{eqnarray}
Let us assume $\lambda^{\alpha}=a^{p-1}$.
Then the relations (\ref{scaling2}) after simple evaluations take the following form:
\begin{eqnarray}
	\frac{M_{F}(X)}{M_{0}}a^{n}=\frac{(a)^{p\,(X+\theta)+\chi}-(a)^{p\,(-X-\theta)-\chi}}{(a)^{p\,(X+\theta)+\chi}+(a)^{p\,(-X-\theta)-\chi}},\nonumber\\
	\frac{M_{G}(X)}{M_{0}}a^{n}=\frac{(a)^{p\,(X-\theta)+\chi}-(a)^{p\,(-X+\theta)-\chi}}{(a)^{p\,(X-\theta)+\chi}+(a)^{p\,(-X+\theta)-\chi}},\label{scaling3}
\end{eqnarray}
where $n=\frac{\nu}{\alpha}(p-1)$ and $p\in \{\mathbf{R}/0\}$.
 Now we are ready to introduce $f,B_{m}$ and $T$ as independent variables of the considered model. Let us assume that the base $a$ is a function of these independent variables:
\begin{equation}
\label{parametr1}
a=a(f,B_{m},T),
\end{equation}
which satisfy assumptions for the homogeneous function in general sense. Let there exist $\{\sigma,\beta,\gamma,\delta\}\in \mathbf{R}^{4}$ such that $\forall{\tilde{\lambda}} \in \mathbf{R}_{+}$ the following relation holds: 
\begin{equation}
\label{scaling4}
{\tilde{\lambda}}^{\sigma}\,a(f,B_{m},T)=a({\tilde{\lambda}}^{\beta}f,{\tilde{\lambda}}^{\gamma}B_{m},{\tilde{\lambda}}^{\delta}T) 
\end{equation}

 Basing on experience of many researches in applying the scaling, we assume that the following substitution for ${\tilde{\lambda}}$:   

\begin{equation}
\label{scaling2p}
{\tilde{\lambda}}=B_{m}^{-1/\gamma}. 
\end{equation}
Then the expression for $a(f,B_{m},T)$ after simple evaluations takes the following form:
\begin{equation}
\label{scaling3p}
a(f,B_{m},T)=B_{m}^{\zeta}\, b(B_{m}^{-\eta}f,B_{m}^{-\rho}T),
\end{equation}
where, $b(\cdot,\cdot)$ is an arbitrary function of the two variables,
\begin{equation}
\label{greek}
 \zeta=\frac{\sigma}{\gamma},\eta=\frac{\beta}{\gamma},\rho=\frac{\delta}{\gamma}. 
\end{equation}
 All parameters in (\ref{scaling3}) $n,M_{0},p,\theta,\chi$ and ones in (\ref{scaling3p}) $b(\cdot,\cdot),\zeta,\eta,$ and $\rho$ must be determined from an experimental data. By introducing new transformstion represented by ${\tilde{\lambda}}$, we have extent the symmetry group
 of the loops space to the three parameters one: 
 \begin{equation}
 \label{lkl}
\mathcal{G}_{\lambda,\chi,\tilde{\lambda}}=(\mathcal{G}_{\lambda} \rtimes \mathcal{G}_{\chi})\times\mathcal{G}_{\tilde{\lambda}}. 
 \end{equation}

\subsection{Modeling of b$(\cdot,\cdot)$ function} 
In this subsection we present an example model of the $b(B_{m}^{-\frac{\beta}{\gamma}}f,B_{m}^{-\frac{\delta}{\gamma}}T)$ function. Applying the idea presented in \citep{bib:Rusz} we factorize
the considered function in the following way:
\begin{equation}
\label{factor}
 b(B_{m}^{-\eta}f,B_{m}^{-\rho}T)=F(B_{m}^{-\eta}f)\Theta(B_{m}^{-\rho}T),
\end{equation}
where $F(\cdot)$ and $\Theta(\cdot)$ have to be model. For $F(\cdot)$ we select a power series, whereas for $\Theta(\cdot)$ the Pad\'{e} approximant has been chosen \citep{bib:pade}
%\begin{figure}%[!t]
%\begin{center}
%\includegraphics[ width=10cm]{Graphmultiloopchi.eps}
%\caption{Magnetic hysteresis family for %$p=1\hspace{2mm}n=1\hspace{2mm}\theta=1,3\hspace{2mm}\nu/\alpha=1$.}
%\label{Fig.4}
%\end{center}
%\end{figure} 
\begin{equation}
\label{Mac}
F\left(\frac{f}{B_{m}^{\eta}}\right)=\Gamma_{0}+\Gamma_{1}\frac{f}{B_{m}^{\eta}}+\dots\Gamma_{k}\left(\frac{f}{B_{m}^{\eta}}\right)^{k}
\end{equation}

Basing on  some computer experiments we have selected for $\Theta(\cdot)$ the following Pad\'{e} approximant \citep{bib:pade}:
\begin{equation}
	\label{scal4}
	\Theta=\left(\frac{\psi_{0}+\theta\,(\psi_{1}+\theta\,\psi_{2})}{1+\theta\,(\psi_{3}+\theta\,\psi_{4})}\right)^{1-z},
\end{equation}
where $\theta=\frac{T+\tau}{ B_{m}^{\rho}}$ and $T$ is measured temperature in $^{\circ}$C.   $\tau$ and $z$ are tuning parameters, $\psi_{i}$ are Pad\'{e} expansion coefficients.

 \section{Conclusions}\label{konk}
In this paper we have widen the simple model of magnetic hysteresis on the set of the following features: frequency, pick of induction and temperature dependence. By generalization of the base $a$ to the homogenous function of the three variables we considered the two cases. First, we have opened new possibility for description of influence of frequency and temperature on the hysteresis phenomena. However, this achievement is conditional. We have assumed that $a(f,B_{m},T)$ is a homogeneous function in general sense. But, this need not to be so. However, if in an individual case, one prove by using experimental data that $a(f,B_{m},T)$ is a homogenous function \citep{bib:SOK1}, then the scaling (\ref{scaling4}) holds and this can be applied in modeling. In the second case we considered $a=const$.   As we have shown in the case each value of $a>0$ constitutes orbit in $\mathcal{G}_{\lambda,\chi}$ defined by semi-direct product  of multiplicative and additive subgroups: $\mathcal{G_{\lambda}}$ and $\mathcal{G_{\chi}}$,respectively. Each element of $\mathcal{G}_{\lambda,\chi}$ relates the two loops. This relation is equivalence one  and introduces division of the loops’ space. However, if an extension of $a$ to a function $a(f,B_{m},T)$ obeys the scaling, then  each two bases can be related by an element of $\mathcal{G}_{\lambda,\chi,\tilde{\lambda}}$. Therefore, in this case all loops are equivalent and all of them constitute one orbit. Decreasing symmetry from $\mathcal{G}_{\lambda,\chi,\tilde{\lambda}}$ to $\mathcal{G}_{\lambda,\chi}$ we reveal splitting of the one lops' set into  infinite number of subsets.

\bibliographystyle{plainnat}

\begin{thebibliography}{99}
\bibitem{bib:sok0}
K.Z. Sokalski, {\em Acta Phys. Pol. series a}, \textbf{127}, 850 (2015).
\bibitem{bib:JILES} 
D.C. Jiles, Frequency dependence of hysteresis curves in conducting magnetic materials, Journal of Applied Physics \textbf{76}, 5849 (1994).
\bibitem{bib:sutor}
A. Sutor, S.J. Rupitsch, S. Bi, and R. Lerch, A modified Preisach hysteresis operator for the modeling of temperature dependent magnetic material behavior, Journal of Applied Physics, \textbf{109}, 07D338 (2011); doi: 10.1063/1.3562520.
\bibitem{bib:BAR}
G.I Barenblatt, Scaling, Cambridge Texts in Applied Mathematics, Cambridge University Press 2003.
\bibitem{bib:SOK1}
K.Z. Sokalski, J. Szczyg{\l}owski, M. Najgebauer and W. Wilczy\,{n}ski, Losses scaling in soft
magnetic materials, COMPEL: Int. J. Comput. Math. Electr. Electron. Eng.,\textbf{26},640-649( 2007).
\bibitem{bib:Rusz}
A. Ruszczyk, K.Z. Sokalski, COMPEL: Int. J. Comput. Math. Electr. Electron. Eng.,\textbf{34},371-379( 2015).
\bibitem{bib:reason}
J. Cale, S. D. Sudhoff, and R. R. Chan, A Field-Extrema Hysteresis Loss Model for High-Frequency
Ferrimagnetic Materials, IEEE TRANSACTIONS ON MAGNETICS, \textbf{44}, 1728-1736 (2008).

\bibitem{bib:pade}
William H. Press, Saul A. Teukolsky, William T. Vetterling, Brian P. Flannery, Numerical Recipes in Fortran 77, The Art of Scientific Computing, Second Edition, Volume 1 of Fortran Numerical Recipes, Published by the Press Syndicate of the University of Cambridge 1997, p. 194.
%\bibitem{bib:Hamm}
%M. Hamermesh, Group Theory and Its Application to Physical Problems, Pergamon Press, 1962. 
%\bibitem{bib:Pre}
%F. Preisach, {\"U}ber die magnetische Nachwirkung, Z. Phys. \textbf{94} 277-302 (1935).
%\bibitem{bib:SLUS}
%B \'{S}lusarek, B. Jankowski, K.  Sokalski, J. Szczyg{\l}owski, Characteristics of power loss in soft magnetic composites a key
%for designing the best values of technological parameters, J. Alloys. Comp. \textbf{581} 699-704 (2013).
%\bibitem{bib:SOKNEW}
%K.Z. Sokalski, B. Jankowski, B. \,{S}lusarek, Binary relations between magnitudes of different dimensions used in material science optimization problems. Pseudo-state equation of Soft Magnetic Composites, Materials Sciences and Applications, 12/2014; 5(12A 36)
\end{thebibliography}
 
\end{document}